\documentclass[a4paper,twocolumn,10pt]{quantumarticle}
\pdfoutput=1

\usepackage{xcolor}
\usepackage[pdftex]{graphicx}
\usepackage{microtype}
\usepackage{hyperref}
\usepackage[braket, qm]{qcircuit}
\usepackage{graphicx}
\usepackage[cmex10]{amsmath}
\usepackage{array}
\usepackage{url}
\usepackage{braket}
\usepackage{amssymb}
\usepackage{caption}
\usepackage{tikz}
\usepackage{adjustbox}
\usepackage{subcaption}
\usepackage{tcolorbox}
\usepackage[frozencache,cachedir=.]{minted}
\setminted[python]{
fontsize = \footnotesize,
}

\tcbset{colback=black!5!white,arc=2pt,boxrule=0.5pt}
\BeforeBeginEnvironment{minted}{\begin{tcolorbox}}
\AfterEndEnvironment{minted}{\end{tcolorbox}}

\newcommand{\ttfnt}[1]{{\small \ttfamily #1}}

\begin{document}

\title{Qrisp: A Framework for Compilable High-Level Programming of Gate-Based Quantum Computers}

\author{Raphael Seidel}
\email{raphael.seidel@fokus.fraunhofer.de}
\affiliation{Fraunhofer Institute for Open Communication Systems, Berlin, Germany}

\author{Sebastian Bock}
\email{[firstname].[lastname]@fokus.fraunhofer.de}
\affiliation{Fraunhofer Institute for Open Communication Systems, Berlin, Germany}

\author{René Zander}
\affiliation{Fraunhofer Institute for Open Communication Systems, Berlin, Germany}

\author{Matic Petri\v{c}}
\affiliation{Fraunhofer Institute for Open Communication Systems, Berlin, Germany}

\author{Niklas Steinmann}
\affiliation{Fraunhofer Institute for Open Communication Systems, Berlin, Germany}

\author{Nikolay Tcholtchev}
\affiliation{Fraunhofer Institute for Open Communication Systems, Berlin, Germany}
\author{Manfred Hauswirth}
\affiliation{Fraunhofer Institute for Open Communication Systems, Berlin, Germany}
\affiliation{Technische Universität Berlin, Berlin, Germany}

\maketitle
\vspace{-0.3em}
\begin{abstract}
While significant progress has been made on the hardware side of quantum computing, support for high-level quantum programming abstractions remains underdeveloped compared to classical programming languages. In this article, we introduce \textit{Qrisp}, a framework designed to bridge several gaps between high-level programming paradigms in state-of-the-art software engineering and the physical reality of today's quantum hardware. The framework aims to provide a systematic approach to quantum algorithm development such that they can be effortlessly implemented, maintained and improved. We propose a number of programming abstractions that are inspired by classical paradigms, yet consistently focus on the particular needs of a quantum  developer. Unlike many other high-level language approaches, \textit{Qrisp}'s standout feature is its ability to compile programs to the circuit level, making them executable on most existing physical backends. The introduced abstractions enable the \textit{Qrisp} compiler to leverage algorithm structure for increased compilation efficiency. Finally, we present a set of code examples, including an implementation of Shor's factoring algorithm. For the latter, the resulting circuit shows significantly reduced quantum resource requirements, strongly supporting the claim that systematic quantum algorithm development can give quantitative benefits.
\end{abstract} 

\vspace{-0.5em}
\section{Overview}
Finding the ``right'' abstractions to best support quantum programmers and enable traditional software engineering features such as efficient debugging, meaningful encapsulation of reusable code, cross-platform development and high-level programming abstractions are still under active research. Compared to traditional programming, quantum programming is still at an ``assembler level''. Over the last 20 years programming a real quantum computer has had more or less the same simplistic structure and basically means adding gates to a quantum circuit by hand. This was feasible for the Qubit scales of the past, but already for current systems with more than 100 Qubits, writing complex programs can quickly turn into a challenging problem. As we know from software engineering research, complexity leads to increases in bugs, poor code maintainability, and consequently to higher costs of software development. Future quantum computers with 1000 and more Qubits will make these shortcomings even more apparent.

It is therefore of paramount importance that quantum computers eventually can be controlled and programmed using high level languages, shifting the focus from the nitty-gritty hardware details to modern software engineering and concepts.
The problem of lacking high-level abstractions for programming a quantum computer has been recognized by the research community and a number of higher-level languages have been proposed. However, many of them either still address Qubits and add gates manually \cite{q_sharp, scaffold, quipper}, or are of a theoretical nature \cite{silq, Qunity}, as they provide no means of compiling/connecting a program to actual quantum hardware as of now. 

\textit{Qrisp} bridges several of the existing gaps and provides a framework with the following design goals:
\begin{enumerate}
\item A programming interface, which seamlessly connects multiple layers of abstraction to facilitate scalable development and compilation of a diverse set of quantum algorithms.
\item A compiler architecture that enables hardware-specific compilation of hardware-agnostic code.
\item An educational tool to introduce quantum computing from top to bottom, rather than starting at the circuit level.
\end{enumerate}
Throughout this paper we present multiple instances of programming paradigms that enable a highly efficient compilation. Despite the fact, that \textit{Qrisp} performs barely any low-level optimization, as it is common for other state-of-the-art frameworks like \cite{pyzx, Qiskit, tket, cirq_developers_2022_6599601}, \textit{Qrisp} facilitates significant compilation improvements as the compiler can exploit knowledge of the high-level algorithm structure.\\

As an embedded domain specific language (eDSL), \textit{Qrisp} is written in Python, as is the source code for \textit{Qrisp}. This allows developers direct access to the vast ecosystem of scientific and industrial libraries that Python has to offer. The framework can create circuit representations (such as OpenQASM \cite{Cross2022}), implying \textit{Qrisp} programs can be run on many of today's physical quantum backends and simulators.\\

This paper provides an overview of the main concepts and features of \textit{Qrisp} along with practical examples. The paper is organized as follows:
\begin{itemize}
    \item Section~\ref{introduction} introduces the core concepts of Qrisp using introductory examples.
    \item Section~\ref{high_level_user_interface} provides an overview of the introduced programming abstractions.
    \item Section~\ref{infrastructure} elaborates on the circuit level manipulation features.
    \item Section~\ref{examples} gives several examples how \textit{Qrisp} code looks in practice.
    \item Section \ref{outlook} points out some of the current challenges of Quantum SDKs and how they could be overcome.
    \item  And Section~\ref{conclusion} summarizes the article and we draw our conclusions.
\end{itemize}

\section{Introduction}
\label{introduction}
The central abstraction of \textit{Qrisp} is the \ttfnt{QuantumVariable}. A \ttfnt{QuantumVariable} hides the qubit management from the user, enabling human readable inputs and outputs, strong typing via class inheritance, infix arithmetic syntax, and much more. Creating a \ttfnt{QuantumVariable} is simple:
\begin{minted}{python}
from qrisp import QuantumVariable
qv = QuantumVariable(size=5)
\end{minted}
Here, $size = 5$ refers to the number of qubits the \ttfnt{QuantumVariable} represents. In the following we show small code snippets to illustrate how a \ttfnt{QuantumVariable} can be used. First, we create a \ttfnt{QuantumFloat}, which is a subclass of \ttfnt{QuantumVariable}:

\begin{minted}{python}
test_float = QuantumFloat(msize=3,
                          exponent=-2,
                          signed=False)
\end{minted}
This statement creates a \ttfnt{QuantumFloat} with 3 mantissa qubits and an exponent of -2, implying the value range has a precision of $2^{-2}$. The values this float can represent are therefore $\{0.0, 0.25, 0.5, 0.75, 1.0, 1.25, 1.5, 1.75\}$. The expression \ttfnt{signed = False} implies that the represented values carry no sign. This keyword argument is optional - by default the constructor creates unsigned \ttfnt{QuantumFloat} variables. For the encoding of the values, we refer to the documentation of our arithmetic module \cite{seidel2021efficient}.
To encode a value into the \ttfnt{QuantumFloat} variable, we use the slicing operator symbol.

\begin{minted}{python}
test_float[:] = 0.5
\end{minted}
The \ttfnt{QuantumFloat} class allows us to use arithmetic operations just as we are used to.
As an example we create a second \ttfnt{QuantumFloat} and multiply both:

\begin{minted}{python}
factor = QuantumFloat(3, 0)
factor[:] = 3

result = factor*test_float
print(result)
\end{minted}
Note that the multiplication quantum algorithm behind such an expression can be interchanged effortlessly, which facilitates hardware specific compilation of hardware agnostic high-level code. In bra-ket notation this program can be written as
\begin{align}
\begin{aligned}
U_{\text{mult}} U_{\text{encode}} \ket{0} \ket{0} \ket{0} &= U_{\text{mult}} \ket{0.5} \ket{3} \ket{0}\\
&= \ket{0.5} \ket{3} \ket{0.5\cdot3}.
\end{aligned}
\end{align}
The console output then reads:
\begin{align*}
\{1.5:\  1.0\}.
\end{align*}
The \ttfnt{print} statement invokes a simulator, which evaluates the generated circuit. The results are returned in the form of a dictionary of bit-strings and are then converted into the corresponding outcome label. This label is determined by the \ttfnt{decoder} method of the \ttfnt{QuantumVariable}.\\
Here, the number $1.5$ represents the label of the measured state, i.e. \ttfnt{3*0.5}, and $1.0$ is the probability of measuring that state.
In order to introduce some quantumness into the program, we add a third \ttfnt{QuantumFloat}, but this time in a state of superposition:
\begin{minted}{python}
from qrisp import h

summand = QuantumFloat(3, -1)
summand[:] = 2
h(summand[0])

result += summand
print(result)
\end{minted}
After encoding the value 2 into \ttfnt{summand}, we apply a Hadmard gate onto the 0-th qubit. This qubit corresponds to the least significant digit of \ttfnt{summand}, which is in this case 0.5, since we are dealing with an exponent of -1. Therefore, this brings \ttfnt{summand} into the superposition state 
\begin{align}
\ket{\text{\ttfnt{summand}}} = \frac{1}{\sqrt{2}} \left( \ket{2} + \ket{2.5} \right).
\end{align}
After executing the \ttfnt{print} statement, the console reads
\begin{align*}
\{3.5: 0.5,\  4: 0.5\}
\end{align*}
which corresponds to the expected outcome.

Arbitrary \textit{Qrisp} code can be exported to a \ttfnt{QuantumCircuit} representation (see Section \ref{infrastructure}) but also be run on IBM backends:

\begin{minted}{python}
from qiskit_ibm_provider import IBMProvider
prv = IBMProvider(YOUR_APITOKEN)
kolkata_qsk = prv.get_backend("ibm_kolkata")
from qrisp import VirtualQiskitBackend
kolkata_qrisp = VirtualQiskitBackend(kolkata_qsk)
\end{minted}
The object \ttfnt{kolkata\_qrisp} is now a \textit{Qrisp} backend object, that can be used to evaluate Qrisp programs. We can call the \ttfnt{get\_measurement} method of the \ttfnt{result} variable of the previous script:
\begin{minted}{python}
meas_res = result.get_measurement(
           backend = kolkata_qrisp)
\end{minted}

\subsection{Solving a quadratic equation}
\label{qdr_eq}

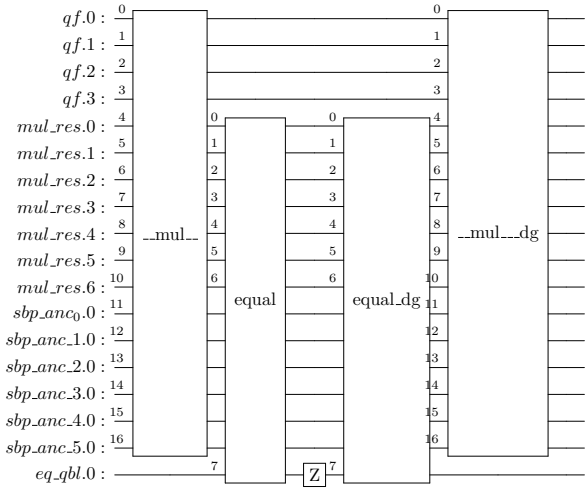
\begin{figure}
\hspace{-1em}
\scalebox{0.68}{
\Qcircuit @C=1.0em @R=0.2em @!R { \\
	 	\nghost{{qf.0} :  } & \lstick{{qf.0} :  } & \multigate{16}{\mathrm{\_\_mul\_\_}}_<<<{0} & \qw & \qw & \qw & \multigate{16}{\mathrm{\_\_mul\_\_\_dg}}_<<<{0} & \qw & \qw\\
	 	\nghost{{qf.1} :  } & \lstick{{qf.1} :  } & \ghost{\mathrm{\_\_mul\_\_}}_<<<{1} & \qw & \qw & \qw & \ghost{\mathrm{\_\_mul\_\_\_dg}}_<<<{1} & \qw & \qw\\
	 	\nghost{{qf.2} :  } & \lstick{{qf.2} :  } & \ghost{\mathrm{\_\_mul\_\_}}_<<<{2} & \qw & \qw & \qw & \ghost{\mathrm{\_\_mul\_\_\_dg}}_<<<{2} & \qw & \qw\\
	 	\nghost{{qf.3} :  } & \lstick{{qf.3} :  } & \ghost{\mathrm{\_\_mul\_\_}}_<<<{3} & \qw & \qw & \qw & \ghost{\mathrm{\_\_mul\_\_\_dg}}_<<<{3} & \qw & \qw\\
	 	\nghost{{mul\_res.0} :  } & \lstick{{mul\_res.0} :  } & \ghost{\mathrm{\_\_mul\_\_}}_<<<{4} & \multigate{13}{\mathrm{equal}}_<<<{0} & \qw & \multigate{13}{\mathrm{equal\_dg}}_<<<{0} & \ghost{\mathrm{\_\_mul\_\_\_dg}}_<<<{4} & \qw & \qw\\
	 	\nghost{{mul\_res.1} :  } & \lstick{{mul\_res.1} :  } & \ghost{\mathrm{\_\_mul\_\_}}_<<<{5} & \ghost{\mathrm{equal}}_<<<{1} & \qw & \ghost{\mathrm{equal\_dg}}_<<<{1} & \ghost{\mathrm{\_\_mul\_\_\_dg}}_<<<{5} & \qw & \qw\\
	 	\nghost{{mul\_res.2} :  } & \lstick{{mul\_res.2} :  } & \ghost{\mathrm{\_\_mul\_\_}}_<<<{6} & \ghost{\mathrm{equal}}_<<<{2} & \qw & \ghost{\mathrm{equal\_dg}}_<<<{2} & \ghost{\mathrm{\_\_mul\_\_\_dg}}_<<<{6} & \qw & \qw\\
	 	\nghost{{mul\_res.3} :  } & \lstick{{mul\_res.3} :  } & \ghost{\mathrm{\_\_mul\_\_}}_<<<{7} & \ghost{\mathrm{equal}}_<<<{3} & \qw & \ghost{\mathrm{equal\_dg}}_<<<{3} & \ghost{\mathrm{\_\_mul\_\_\_dg}}_<<<{7} & \qw & \qw\\
	 	\nghost{{mul\_res.4} :  } & \lstick{{mul\_res.4} :  } & \ghost{\mathrm{\_\_mul\_\_}}_<<<{8} & \ghost{\mathrm{equal}}_<<<{4} & \qw & \ghost{\mathrm{equal\_dg}}_<<<{4} & \ghost{\mathrm{\_\_mul\_\_\_dg}}_<<<{8} & \qw & \qw\\
	 	\nghost{{mul\_res.5} :  } & \lstick{{mul\_res.5} :  } & \ghost{\mathrm{\_\_mul\_\_}}_<<<{9} & \ghost{\mathrm{equal}}_<<<{5} & \qw & \ghost{\mathrm{equal\_dg}}_<<<{5} & \ghost{\mathrm{\_\_mul\_\_\_dg}}_<<<{9} & \qw & \qw\\
	 	\nghost{{mul\_res.6} :  } & \lstick{{mul\_res.6} :  } & \ghost{\mathrm{\_\_mul\_\_}}_<<<<{10} & \ghost{\mathrm{equal}}_<<<{6} & \qw & \ghost{\mathrm{equal\_dg}}_<<<{6} & \ghost{\mathrm{\_\_mul\_\_\_dg}}_<<<<{10} & \qw & \qw\\
	 	\nghost{{sbp\_anc_0.0} :  } & \lstick{{sbp\_anc_0.0} :  } & \ghost{\mathrm{\_\_mul\_\_}}_<<<<{11} & \ghost{\mathrm{equal}} & \qw & \ghost{\mathrm{equal\_dg}} & \ghost{\mathrm{\_\_mul\_\_\_dg}}_<<<<{11} & \qw & \qw\\
	 	\nghost{{sbp\_anc\_1.0} :  } & \lstick{{sbp\_anc\_1.0} :  } & \ghost{\mathrm{\_\_mul\_\_}}_<<<<{12} & \ghost{\mathrm{equal}} & \qw & \ghost{\mathrm{equal\_dg}} & \ghost{\mathrm{\_\_mul\_\_\_dg}}_<<<<{12} & \qw & \qw\\
	 	\nghost{{sbp\_anc\_2.0} :  } & \lstick{{sbp\_anc\_2.0} :  } & \ghost{\mathrm{\_\_mul\_\_}}_<<<<{13} & \ghost{\mathrm{equal}} & \qw & \ghost{\mathrm{equal\_dg}} & \ghost{\mathrm{\_\_mul\_\_\_dg}}_<<<<{13} & \qw & \qw\\
	 	\nghost{{sbp\_anc\_3.0} :  } & \lstick{{sbp\_anc\_3.0} :  } & \ghost{\mathrm{\_\_mul\_\_}}_<<<<{14} & \ghost{\mathrm{equal}} & \qw & \ghost{\mathrm{equal\_dg}} & \ghost{\mathrm{\_\_mul\_\_\_dg}}_<<<<{14} & \qw & \qw\\
	 	\nghost{{sbp\_anc\_4.0} :  } & \lstick{{sbp\_anc\_4.0} :  } & \ghost{\mathrm{\_\_mul\_\_}}_<<<<{15} & \ghost{\mathrm{equal}} & \qw & \ghost{\mathrm{equal\_dg}} & \ghost{\mathrm{\_\_mul\_\_\_dg}}_<<<<{15} & \qw & \qw\\
	 	\nghost{{sbp\_anc\_5.0} :  } & \lstick{{sbp\_anc\_5.0} :  } & \ghost{\mathrm{\_\_mul\_\_}}_<<<<{16} & \ghost{\mathrm{equal}} & \qw & \ghost{\mathrm{equal\_dg}} & \ghost{\mathrm{\_\_mul\_\_\_dg}}_<<<<{16} & \qw & \qw\\
	 	\nghost{{eq\_qbl.0} :  } & \lstick{{eq\_qbl.0} :  } & \qw & \ghost{\mathrm{equal}}_<<<{7} & \gate{\mathrm{Z}} & \ghost{\mathrm{equal\_dg}}_<<<{7} & \qw & \qw & \qw\\
\\ }}
\caption{\label{sqrt_oracle_circ} The circuit generated by calling \ttfnt{sqrt\_oracle} on \ttfnt{qf}. Note that the qubits named \ttfnt{sbp\_anc} are ancillae utilized by the multiplication algorithm. These ancillae are successively computed and uncomputed, such that after compilation (invoked via \ttfnt{qf.qs.compile()}) they are allocated and deallocated on the same physical qubit, which will (temporarily) also hold the value of \ttfnt{eq\_qbl}. The quantum circuit resulting from the \ttfnt{compile} method therefore contains 12 qubits.}
\end{figure}

Moving on from trivial expressions, we provide an example of solving the quadratic equation
\begin{align}
    x^2 = 0.25
\end{align}
using Grover's algorithm \cite{grover}. The idea here is to prepare an oracle, that multiplies a \ttfnt{QuantumFloat} with itself and tags the desired value $c_{tag} = 0.25$. This oracle is then embedded into several Grover iterations to amplify the amplitude of the solution.
\subsubsection{Oracle Construction}
The first step is to construct the oracle.
\begin{minted}{python}
from qrisp import auto_uncompute, QuantumFloat, z

@auto_uncompute
def sqrt_oracle(qf):
    sqr_res = qf*qf
    comparison_res = (sqr_res == 0.25)
    z(comparison_res)
\end{minted}
It is important to note that the oracle is not a framework specific gate object as it would be the case for frameworks like Cirq~\cite{cirq_developers_2022_6599601} or Qiskit~\cite{Qiskit}, but instead a Python function taking a \ttfnt{QuantumVariable} as parameter. This relieves the user from specifying which qubits are supposed to be ``plugged'' into which input of the oracle gate object, and facilitates direct interoperability with \ttfnt{QuantumFloat}s of different size, exponent, and signed-status. Perhaps the biggest advantage of creating quantum algorithms using \ttfnt{QuantumVariable}s is the potential for modularity: Since the allocation of ``algorithmic'' qubits to physical qubits is both abstracted and automated, developers working on different modules no longer need to communicate in order to be able to reuse recycled qubits from each others module. Instead, they just call the \ttfnt{QuantumVariable} constructor, which will find the best options out of all previously deallocated qubits.

The first operation in the above definition of \ttfnt{sqrt\_oracle}  multiplies the quantum variable provided as parameter with itself and stores the result of the multiplication into a newly created \ttfnt{QuantumFloat} called \ttfnt{sqr\_res}.
The next step is to compare to the target value $c_{tag} = 0.25$ using the \ttfnt{==} operator as known from classical computing. This comparison returns a \ttfnt{QuantumBool} holding the comparisons result (possibly in superposition). Finally, we apply a Z-gate on \ttfnt{comparison\_res} to apply the phase-flip to the states satisfying $x^2 = 0.25$.

An important detail of the code snippet above is the \ttfnt{auto\_uncompute} decorator in line 3. This decorator ensures that all local \ttfnt{QuantumVariables} of this function, i.e., \ttfnt{sqr\_res} and \ttfnt{comparison\_res}, are disentangled/uncomputed properly after the function \ttfnt{sqrt\_oracle} is executed. This is important for qubit resource management: The uncomputed qubits will be  reused automatically for other purposes, but also because the Grover diffuser needs its operand to be disentangled in order to work as intended.
The underlying algorithm for synthesizing the uncomputed quantum circuit has been proposed by Paradis et al. \cite{unqomp} and has been refined in \cite{seidel_2023_uncomp}.

A representation of the resulting circuit after applying \ttfnt{sqrt\_oracle} to a \ttfnt{QuantumFloat} is provided in Fig.~\ref{sqrt_oracle_circ}.

\subsubsection{Grover's algorithm}
\label{grovers_alg}
Now we embed the constructed oracle into Grover's algorithm.

\begin{minted}{python}
from qrisp import QuantumFloat, h
from qrisp.grover import diffuser

qf = QuantumFloat(3, -1, signed = True)

n = qf.size 
iterations = int((2**n/2)**0.5)

h(qf)

for i in range(iterations):
    sqrt_oracle(qf)
    diffuser(qf)

result = qf.get_measurement(plot = True)
\end{minted}

\begin{figure*}
\begin{subfigure}{.55\textwidth}
\scalebox{0.68}{
\Qcircuit @C=1.0em @R=0.2em @!R { \\
	 	\nghost{{qf.0} :  } & \lstick{{qf.0} :  } & \gate{\mathrm{H}} & \multigate{17}{\mathrm{sqrt\_oracle}}_<<<{0} & \multigate{3}{\mathrm{diffuser}}_<<<{0} & \multigate{17}{\mathrm{sqrt\_oracle}}_<<<{0} & \multigate{3}{\mathrm{diffuser}}_<<<{0} & \qw & \qw\\
	 	\nghost{{qf.1} :  } & \lstick{{qf.1} :  } & \gate{\mathrm{H}} & \ghost{\mathrm{sqrt\_oracle}}_<<<{1} & \ghost{\mathrm{diffuser}}_<<<{1} & \ghost{\mathrm{sqrt\_oracle}}_<<<{1} & \ghost{\mathrm{diffuser}}_<<<{1} & \qw & \qw\\
	 	\nghost{{qf.2} :  } & \lstick{{qf.2} :  } & \gate{\mathrm{H}} & \ghost{\mathrm{sqrt\_oracle}}_<<<{2} & \ghost{\mathrm{diffuser}}_<<<{2} & \ghost{\mathrm{sqrt\_oracle}}_<<<{2} & \ghost{\mathrm{diffuser}}_<<<{2} & \qw & \qw\\
	 	\nghost{{qf.3} :  } & \lstick{{qf.3} :  } & \gate{\mathrm{H}} & \ghost{\mathrm{sqrt\_oracle}}_<<<{3} & \ghost{\mathrm{diffuser}}_<<<{3} & \ghost{\mathrm{sqrt\_oracle}}_<<<{3} & \ghost{\mathrm{diffuser}}_<<<{3} & \qw & \qw\\
	 	\nghost{{mul\_res.0} :  } & \lstick{{mul\_res.0} :  } & \qw & \ghost{\mathrm{sqrt\_oracle}}_<<<{4} & \qw & \ghost{\mathrm{sqrt\_oracle}}_<<<{4} & \qw & \qw & \qw\\
	 	\nghost{{mul\_res.1} :  } & \lstick{{mul\_res.1} :  } & \qw & \ghost{\mathrm{sqrt\_oracle}}_<<<{5} & \qw & \ghost{\mathrm{sqrt\_oracle}}_<<<{5} & \qw & \qw & \qw\\
	 	\nghost{{mul\_res.2} :  } & \lstick{{mul\_res.2} :  } & \qw & \ghost{\mathrm{sqrt\_oracle}}_<<<{6} & \qw & \ghost{\mathrm{sqrt\_oracle}}_<<<{6} & \qw & \qw & \qw\\
	 	\nghost{{mul\_res.3} :  } & \lstick{{mul\_res.3} :  } & \qw & \ghost{\mathrm{sqrt\_oracle}}_<<<{7} & \qw & \ghost{\mathrm{sqrt\_oracle}}_<<<{7} & \qw & \qw & \qw\\
	 	\nghost{{mul\_res.4} :  } & \lstick{{mul\_res.4} :  } & \qw & \ghost{\mathrm{sqrt\_oracle}}_<<<{8} & \qw & \ghost{\mathrm{sqrt\_oracle}}_<<<{8} & \qw & \qw & \qw\\
	 	\nghost{{mul\_res.5} :  } & \lstick{{mul\_res.5} :  } & \qw & \ghost{\mathrm{sqrt\_oracle}}_<<<{9} & \qw & \ghost{\mathrm{sqrt\_oracle}}_<<<{9} & \qw & \qw & \qw\\
	 	\nghost{{mul\_res.6} :  } & \lstick{{mul\_res.6} :  } & \qw & \ghost{\mathrm{sqrt\_oracle}}_<<<<{10} & \qw & \ghost{\mathrm{sqrt\_oracle}}_<<<<{10} & \qw & \qw & \qw\\
	 	\nghost{{sbp\_anc\_0.0} :  } & \lstick{{sbp\_anc\_0.0} :  } & \qw & \ghost{\mathrm{sqrt\_oracle}}_<<<<{11} & \qw & \ghost{\mathrm{sqrt\_oracle}}_<<<<{11} & \qw & \qw & \qw\\
	 	\nghost{{sbp\_anc\_1.0} :  } & \lstick{{sbp\_anc\_1.0} :  } & \qw & \ghost{\mathrm{sqrt\_oracle}}_<<<<{12} & \qw & \ghost{\mathrm{sqrt\_oracle}}_<<<<{12} & \qw & \qw & \qw\\
	 	\nghost{{sbp\_anc\_2.0} :  } & \lstick{{sbp\_anc\_2.0} :  } & \qw & \ghost{\mathrm{sqrt\_oracle}}_<<<<{13} & \qw & \ghost{\mathrm{sqrt\_oracle}}_<<<<{13} & \qw & \qw & \qw\\
	 	\nghost{{sbp\_anc\_3.0} :  } & \lstick{{sbp\_anc\_3.0} :  } & \qw & \ghost{\mathrm{sqrt\_oracle}}_<<<<{14} & \qw & \ghost{\mathrm{sqrt\_oracle}}_<<<<{14} & \qw & \qw & \qw\\
	 	\nghost{{sbp\_anc\_4.0} :  } & \lstick{{sbp\_anc\_4.0} :  } & \qw & \ghost{\mathrm{sqrt\_oracle}}_<<<<{15} & \qw & \ghost{\mathrm{sqrt\_oracle}}_<<<<{15} & \qw & \qw & \qw\\
	 	\nghost{{sbp\_anc\_5.0} :  } & \lstick{{sbp\_anc\_5.0} :  } & \qw & \ghost{\mathrm{sqrt\_oracle}}_<<<<{16} & \qw & \ghost{\mathrm{sqrt\_oracle}}_<<<<{16} & \qw & \qw & \qw\\
	 	\nghost{{eq\_qbl.0} :  } & \lstick{{eq\_qbl.0} :  } & \qw & \ghost{\mathrm{sqrt\_oracle}}_<<<<{17} & \qw & \ghost{\mathrm{sqrt\_oracle}}_<<<<{17} & \qw & \qw & \qw\\
\\ }}
\caption{\label{grover_circ}}
\end{subfigure}
\begin{subfigure}{.45\textwidth}
    \includegraphics[width = 1\textwidth]{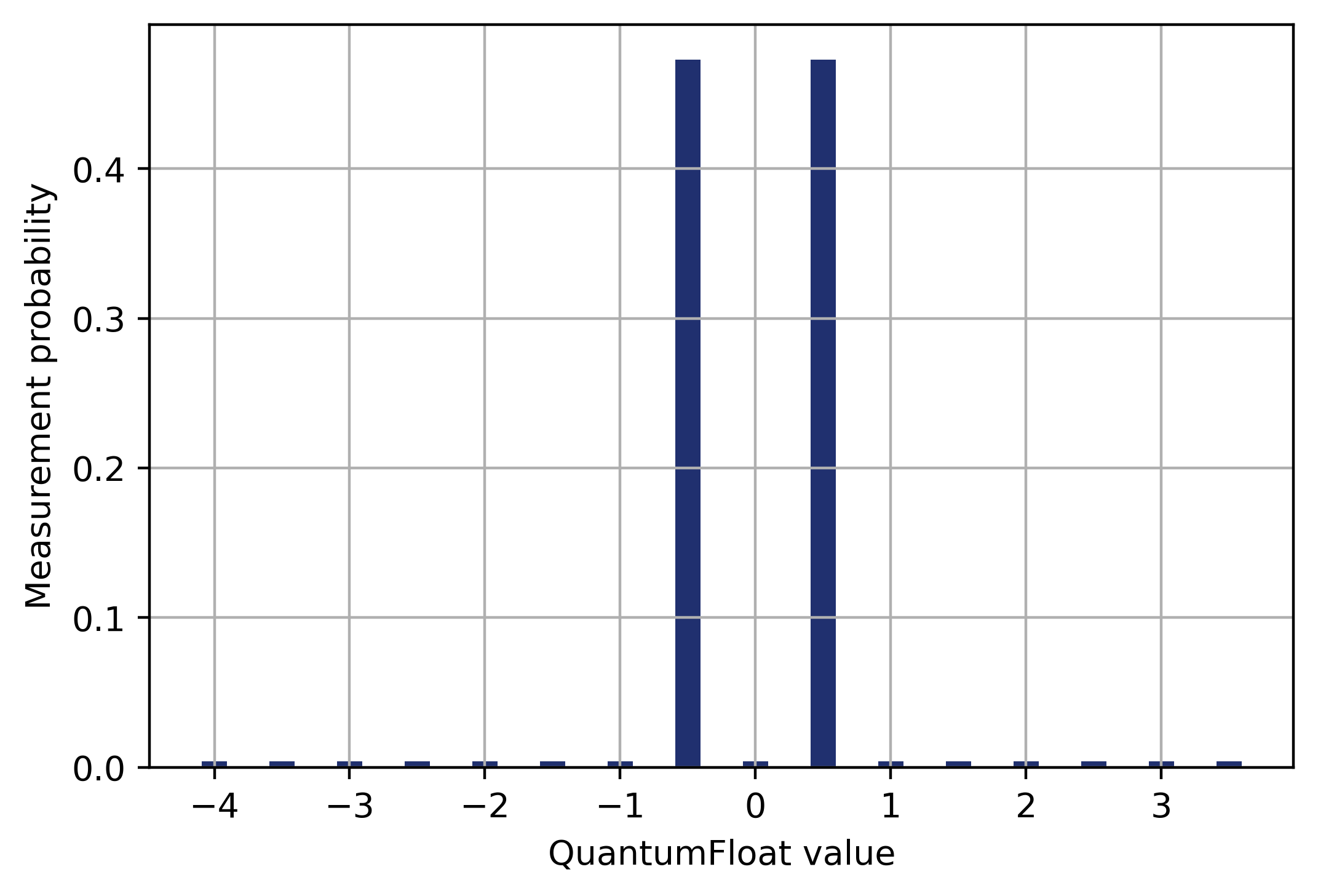}
\caption{\label{mes_plot}}
\end{subfigure}
\caption{\ref{grover_circ} The circuit for Grover's algorithm, which is generated after executing the source code snippet in Section~\ref{grovers_alg}. To improve readability, we hide the circuit displayed in Fig.~\ref{sqrt_oracle_circ} behind the box \ttfnt{sqrt\_oracle}. \ref{mes_plot} A plot of the simulated measurement results of the QuantumFloat \ttfnt{qf}.}
\end{figure*}

First we define a \ttfnt{QuantumFloat}, which will eventually contain the solution.
Then we determine the number of iterations according to the formula given in \cite{grovers_alg_iters}, taking into consideration that we expect two solutions ($S = \{0.5, -0.5\}$). The next step brings \ttfnt{qf} into uniform superposition, followed by the Grover iterations and finalized by a measurement. The result of the measurement and a plot of the generated quantum circuit can be found in Fig.~\ref{mes_plot} and Fig.~\ref{grover_circ}.

\section{High-level Abstractions}
This section provides an overview of some of the core elements that constitute a \textit{Qrisp} program. The following concepts are presented:
\begin{itemize}
    \item \ttfnt{QuantumVariable}s in Section~\ref{quantum_variables}
    \item The \textbf{uncomputation} module in Section~\ref{uncomputation}
    \item The \textbf{typing} system in Section~\ref{quantum_types}
    \item \ttfnt{QuantumSession} and its compilation in Section~\ref{quantum_session}
    \item \ttfnt{QuantumArray} in Section~\ref{quantum_array}
    \item \ttfnt{QuantumDictionary} in Section~\ref{quantum_dictionary}
    \item \ttfnt{QuantumEnvironment} in Section~\ref{quantum_environment}
    \item The \ttfnt{custom\_control} decorator in Section~\ref{custom_control}
\end{itemize}

\label{high_level_user_interface}
\subsection{QuantumVariables}
\label{quantum_variables}
\textit{Qrisp}'s high-level user interface enables the construction of significantly more complex quantum algorithms than it would ever be possible with a gate-based approach. This is achieved by automating many repetitive (and tedious) ``book-keeping'' tasks, such as ancilla qubit management, or measurement outcome decoding. Many of these automations are hidden and performed in the \ttfnt{QuantumVariable} class, which was already discussed in Section~\ref{introduction}. 

\ttfnt{QuantumVariable}s can be manipulated at a low level by calling gate application functions on them:

\begin{minted}{python}
from qrisp import QuantumVariable, z, x, cx
qv = QuantumVariable(5)
x(qv[0])
z(qv)
cx(qv[0], qv[1])
\end{minted}

To perform a measurement, we call the \ttfnt{get\_measurement} method to retrieve a dictionary of the measurement probabilities.

\begin{minted}{python}
meas_res = qv.get_measurement()
print(meas_res)
# Yields {'11000': 1.0}
\end{minted}

As illustrated in Section~\ref{grovers_alg},
once a QuantumVariable is no longer needed, it can be deallocated using the \ttfnt{delete} method. This method should only be called if the register of the \ttfnt{QuantumVariable} is in the $\ket{0}$ state, which can be checked by calling delete with the \ttfnt{verify} keyword:

\begin{minted}{python}
qv.delete(verify = True)
\end{minted}

By default, \ttfnt{verify} is set to \ttfnt{False}, because each verification requires a simulation of the underlying quantum circuit. Especially for algorithms creating and deleting a lot of \ttfnt{QuantumVariable}s this can slow down the compilation speed significantly.
In the code snippet above, \ttfnt{qv} is not in the required state, therefore an exception is raised. To circumvent this, we can either manually perform a sequence of gates, such that \ttfnt{qv} reaches the $\ket{0}$ state, or call the \ttfnt{uncompute} method. This method calls the Unqomp algorithm \cite{unqomp} and subsequently the \ttfnt{delete} method.

\begin{minted}{python}
qv.uncompute()
\end{minted}

Deleting/uncomputing \ttfnt{QuantumVariable}s is an important part of the automated qubit management. The deleted qubits will be reused for allocation of other \ttfnt{QuantumVariable}s. More details are provided in Section~\ref{compilation}.

\subsection{Uncomputation}
\label{uncomputation}
Uncomputation is an important aspect of quantum information processing because it allows for efficient use of quantum resources. In classical computing this can be achieved by deleting information and reusing the deleted bits for other purposes. Deleting (or resetting) a qubit is however not a reversible process and is usually performed by measuring the qubit in question and performing a bitflip based on the outcome. This measurement in turn collapses the superposition of other entangled qubits, which are supposed to be unaffected. In many cases, this collapse interferes with the quantum algorithm, such that the resulting state can no longer be used. In some situations, uncomputation is not only relevant as a way to manage quantum resources, but is also necessary for the quantum algorithm to function properly. One such example is Grover’s algorithm, described in Section~\ref{qdr_eq}, where the Grover diffuser requires its operand to be disentangled from other registers.\\

In many cases, uncomputing a \ttfnt{QuantumVariable} can be achieved by inverting the steps required for the computation. While this seems like a simple recipe, it can be ambiguous to detect what gates contributed to the computation. In any case, it is a tedious amount of extra programming work to be done, which should be automated. Fortunately an algorithm called Unqomp for automatic uncomputation has been developed by Paradis et al.~\cite{unqomp}. An important advantage of Unqomp is, that it does not follow the philosophy of simply reverting the computation. This feature enables the algorithm to skip the “un-uncomputation” of values which would be required to recompute in order to perform the uncomputation.\\

A generalized version of Unqomp \cite{seidel_2023_uncomp} has been implemented within \textit{Qrisp} and can be called in two different ways: The first option is the \ttfnt{auto\_uncompute} decorator, which automatically uncomputes all local QuantumVariables of a function. An example of it's usage can be found in Section~ \ref{grovers_alg}.

The second way of calling uncomputation is the \ttfnt{uncompute} method of the \ttfnt{QuantumVariable} class. We adapt the code from Section~\ref{grovers_alg} to deploy this method instead of the decorator:

\begin{minted}{python}
from qrisp import z

def sqrt_oracle(qf):
    sqr_res = qf*qf
    comparison_res = (sqr_res == 0.25)
    z(comparison_res)
    comparison_res.uncompute()
    sqr_res.uncompute()
\end{minted}

Invoking these ways of uncomputation also calls the \ttfnt{delete} method after. Together these methods facilitate how quantum memory can be managed in practice. A more detailed description about uncomputation in \textit{Qrisp} is available here: \cite{seidel_2023_uncomp}.

\subsection{Quantum types}
\label{quantum_types}
The \ttfnt{QuantumVariable} class is the most abstract of its kind. More specific quantum types can be created using class inheritance enabling a smooth integration of \textit{Qrisp} typing into the Python typing system. \textit{Qrisp} provides four specific built-in types:

\begin{enumerate}
    \item \ttfnt{QuantumFloat}, a type for representing numbers (negative and/or non-integer).
    \item \ttfnt{QuantumBool}, a type for representing boolean values.
    \item \ttfnt{QuantumChar}, a type for representing characters.
    \item \ttfnt{QuantumModulus}, a type for representing modular numbers.
\end{enumerate}

For more information about \ttfnt{QuantumFloat} (especially about the encoding and the underlying arithmetic algorithms) we refer to our article~\cite{seidel2021efficient}.

To illustrate how a custom quantum datatype can be implemented using third party modules, we will now provide an example how a \textit{date} quantum datatype can be realized. For this we inherit from \ttfnt{QuantumVariable}

\begin{minted}{python}
import datetime as dt
from qrisp import QuantumVariable

class QuantumDate(QuantumVariable):

    def __init__(self, size, starting_date):
        self.st_dt = starting_date
        QuantumVariable.__init__(self, size)

    def decoder(self, i):
        return self.st_dt + dt.timedelta(i)
\end{minted}

The relevant part here is the decoder method. This method decides how the measurement results of the \ttfnt{QuantumVariable} are converted to human readable values. The default decoder of the \ttfnt{QuantumVariable} returns bitstrings - in our case we overwrite this specification and instead implement a function that returns a \ttfnt{datetime} object, which represents the desired value. Using this system, developers can easily set up the means to represent arbitrarily complicated data-structures in superposition.

We can now create instances of \ttfnt{QuantumDate}:

\begin{minted}{python}
today = dt.date.today()
tomorrow = today + dt.timedelta(1)
even_later = today + dt.timedelta(4)

qd = QuantumDate(size = 3, starting_date = today)

qd[:] = {today : 1j, 
         tomorrow : 0.5,
         even_later : -0.5}
\end{minted}

As previously mentioned, the slicing operator \ttfnt{[:]} can be used to encode values into \ttfnt{QuantumVariable}s. In this case we do not encode one specific value but a quantum state:

\begin{align}
\begin{aligned}
    \ket{\psi} = \sqrt{\frac{2}{3}}(i &\ket{\text{\ttfnt{today}}}\\
    + 0.5 &\ket{\text{\ttfnt{tomorrow}}} \\
    - 0.5 &\ket{\text{\ttfnt{even\_later}}} )
\end{aligned}
\end{align}

To verify, we can evaluate the state by performing a measurement:

\begin{minted}{python}
print(qd) 
# yields:
# {datetime.date(2024, 6, 16): 0.6667,
#  datetime.date(2024, 6, 17): 0.1667,
#  datetime.date(2024, 6, 20): 0.1667}
\end{minted}

\subsection{QuantumSession}
\label{quantum_session}
Another important component of the high-level interface is the \ttfnt{QuantumSession} class. Each \ttfnt{QuantumVariable} is registered in exactly one \ttfnt{QuantumSession}, which manages its lifetime cycle. If an entangling operation between two \ttfnt{QuantumVariable}s, which are registered in different \ttfnt{QuantumSession}s, is performed, the sessions are merged such that both variables are registered in the same session afterwards. This automation implies that in practise the developer doesn't really have to think about \ttfnt{QuantumSession}s.

Apart from the internal management aspect, the \ttfnt{QuantumSession} includes some convenient interfaces:\\

\subsubsection{Statevector assessment}
\label{statevector_assessment}
The \ttfnt{statevector} method enables a concise display of the current quantum state of the algorithm. Effortless assessment of the statevector is vital for quantum algorithm development, as it enables a fast debugging workflow. Many low-level frameworks address this requirement by providing an array of complex numbers, where each entry corresponds to the amplitude of the corresponding computational basis state. For debugging, this is usually not too helpful because the size of the array grows exponentially with the amount of Qubits and, in many cases, contains a lot of zero entries. To extract any useful information from this array, developers typically have to write additional logic, taking into account many specifics of the assessed quantum circuit.\\
Within \textit{Qrisp} we can take a different approach due to the structure induced by \ttfnt{QuantumVariable}s: \textit{Qrisp} returns a SymPy \cite{sympy} quantum state, which exhibits many convenient features such as compact representation, symbolic investigation/simplification or automatic \LaTeX{} code generation.

\begin{minted}{python}
from qrisp import QuantumFloat, h, cx
a = QuantumFloat(3)
b = QuantumFloat(3)
c = QuantumFloat(3)
h(a[0])
cx(a[0], b[1])
cx(a[0], c[2])
print(a.qs.statevector())
# Yields: sqrt(2)*(|0>**3 + |1>*|2>*|4>)/2
\end{minted}

The low-level approach would yield an array with $2^9 = 512$ entries, a much more opaque representation compared to the \textit{Qrisp} approach.\\

To minimize the amount of redundant information, deleted \ttfnt{QuantumVariable}s are removed from the statevector representation instead of being displayed as $\ket{0}$:

\begin{minted}{python}
b.uncompute()
print(a.qs.statevector())
# Yields: sqrt(2)*(|0>**2 + |1>*|4>)/2
\end{minted}

\subsubsection{Compilation}
\label{compilation}
One of the most central features of \textit{Qrisp} is the \ttfnt{compile} method of the \ttfnt{QuantumSession} class. The purpose of this method is to convert a \ttfnt{QuantumSession} into an optimized \ttfnt{QuantumCircuit}. The \ttfnt{compile} method involves several steps, which decrease the required resources significantly. Their detailed description is out of the scope of this paper and will be discussed in a future paper. For understanding its underlying concepts we briefly describe two important internal steps.

\begin{itemize}
    \item The \textbf{qubit allocation} step ensures that qubits that are required at disjoint time intervals of the algorithm can be allocated on the same physical qubit, thus in many cases heavily reducing the mandatory qubits of the resulting quantum circuit. If presented with an allocation request, the compiler will assess the previously deallocated qubits for the set that is best suited to serve that request. Best suited here means ``qubits, that will be available the earliest in an actual run of the algorithm.'' Here, the compiler includes the possibility to incorporate the gate-speed of a given backend architecture to determine what qubits will be available earliest. To illustrate the power of this technique consider the example from Section~\ref{grovers_alg}. The uncompiled algorithm for solving the quadratic equation involves 32 qubits, however, many of those are required only during a short time interval of the algorithm. The allocation step of Qrip reduces the required number by producing a quantum circuit involving only 12 qubits.\\
    Furthermore, it is possible to grant the compilation method additional \textit{workspace} qubits by calling the compile method with an integer argument, indicating the amount of such qubits. This feature allows developers to dynamically trade circuit depth for qubit count. This can be especially helpful if a backend provides more qubits than initially required by a certain algorithm. The \textit{workspace} keyword then allows the user to compile a circuit with the same semantics but with lower depth.
    \item The \textbf{MCX recompilation} step happens simultaneously with the allocation step. The idea behind MCX recompilation is the observation that multi-controlled X gates can be exponentially less costly if supported by the appropriate amount of (dirty) ancilla qubits \cite{balauca_mcx}. In many situations the compiled algorithm previously released some qubit resources (or has a user-granted workspace), which can be used for exactly this purpose. However, this is not always the case, so simply allocating a fixed amount of extra qubits for any multi-controlled X gate can significantly increase the qubit count of the compilation result. To address this problem, \textit{Qrisp} dynamically generates an MCX implementation at compile time that uses exactly the right amount qubits available at that stage of the compilation. Thus we can harness the speed of ancilla supported MCX implementations without any qubit overhead. Furthermore, this process is fully automated, implying that developers never have to deal with this explicitly. Without MCX recompilation the code from Section~\ref{grovers_alg} compiles to a quantum circuit of depth 2228 involving 1720 CNOT gates and 12 qubits. Upon activating this feature we get a depth of 1106 with 928 CNOT gates and the same amount of qubits, i.e., a reduction of almost 50\%! 
\end{itemize}

A more systematic benchmarking of the presented techniques will follow in a future publication, but the given examples already show, that these techniques can provide significant advantages in compilation efficiency. \textit{Qrisp}'s approach to compilation is fundamentally different from other compilation approaches that perform a circuit-in-circuit-out optimization \cite{pyzx},\cite{tket}, \cite{staq}. In contrast, the \textit{Qrisp} compiler can also make use of high-level information about the algorithm to be compiled, enabling much better optimizations.

\subsection{QuantumArray}
\label{quantum_array}
The \ttfnt{QuantumArray} class streamlines the treatment of structured collections of \ttfnt{QuantumVariable}s of the same type. As a subclass of the NumPy \ttfnt{ndarray}, the \ttfnt{QuantumArray} supports many convenient array manipulation methods. Similar to the NumPy equivalent, creating a \ttfnt{QuantumArray} can be achieved by specifying a shape and a qtype:

\begin{minted}{python}
from qrisp import QuantumArray, QuantumFloat, h
qtype = QuantumFloat(5, -2)
q_array = QuantumArray(qtype = qtype, 
                       shape = (2, 2, 2))
\end{minted}

\ttfnt{QuantumArray}s can be indexed, reshaped, sliced etc. just like regular NumPy arrays:

\begin{minted}{python}
q_array_entry = q_array[0,0,1]
reshaped_q_array = q_array.reshape((2,4))
sliced_q_array = reshaped_q_array[1:,::-1]
\end{minted}

For \ttfnt{QuantumArrays} with type \ttfnt{QuantumFloat}, matrix multiplication is available:

\begin{minted}{python}
q_array_1 = QuantumArray(qtype)
q_array_2 = QuantumArray(qtype)
q_array_1[:] = 2*np.eye(2)
q_array_2[:] = [[1,2],[3,4]]
print(q_array_1 @ q_array_2)
# Yields: {OutcomeArray([[2, 4],
#                        [6, 0]]): 1.0}
\end{minted}

For \ttfnt{QuantumArray}s where the size is an integer power of 2, it is even possible to perform indexing based on an integer \ttfnt{QuantumFloat}:

\begin{minted}{python}
from qrisp import (QuantumBool, 
QuantumArray, QuantumFloat, h)

q_array = QuantumArray(qtype = QuantumBool(), 
                       shape = (4,4))
                       
index_0 = QuantumFloat(2)
index_1 = QuantumFloat(2)

index_0[:] = 2
index_1[:] = 1

h(index_0[0])

with q_array[index_0, index_1] as entry:
    entry.flip()
\end{minted}

\subsection{QuantumDictionary}
\label{quantum_dictionary}
The \ttfnt{QuantumDictionaries} is a data structure that enables loading non-algorithmic\footnote{Non-algorithmic means that the data relation doesn't need to be based on some algorithm such as the set of tuples $\{(x, x^2), x \in \mathbb{N}, x < n\}$, but instead can be arbitrary values.} data relations into a superposition. Internally, \ttfnt{QuantumDicionary} key/value pairs are loaded using a conversion to a truth table representation and subsequent implementation via quantum logic synthesis \cite{soeken2017logic, Meuli2019, Porwik2002, Amy_2019, Seidel_2023}.\\
As an inheritor of the Python dictionary, \ttfnt{QuantumDictionary} has all the functionality we are used to from traditional Python:
\begin{minted}{python}
from qrisp import (QuantumDictionary, 
QuantumVariable, multi_measurement)

return_type = QuantumFloat(4, -2)
qd = QuantumDictionary(return_type = return_type)

qd[1] = 1.5
qd[42] = 3
qd["hello"] = 0.25
qd["world"] = 1.75

print(qd[42])
# Yields: 3
\end{minted}
The core functionality of the \ttfnt{QuantumDictionary} is that it can receive \ttfnt{QuantumVariable}s as keys and return the corresponding values as entangled \ttfnt{QuantumVariable}s.
\begin{align}
    U_{\textit{qd}}\ket{x}\ket{0} = \ket{x}\ket{\text{\textit{qd}}[x]}
\end{align}
This behavior can also be verified in \textit{Qrisp} code:
\begin{minted}{python}
class CustomQV(QuantumVariable):
    def __init__(self):
        QuantumVariable.__init__(self, 2)
    def decoder(self, i):
        labels = [1, 42, "hello", "world"]
        return labels[i]
                                
key_qv = CustomQV()

key_qv[:] = {"hello" : 2**-0.5, 
             "world" : 2**-0.5}

print(key_qv)
# Yields: {'hello': 0.5, 'world': 0.5}
\end{minted}
In this snippet, we create a \ttfnt{QuantumVariable} with a custom quantum type. In the next line, we initiate the state
\begin{align}
    \ket{\psi} = \frac{1}{\sqrt{2}}(\ket{\text{hello}} + \ket{\text{world}})
\end{align}

To load the data from the \ttfnt{QuantumDictionary} into the superposition we dereference \ttfnt{qd} with \ttfnt{key\_qv}.
\begin{minted}{python}
value_qv = qd[key_qv]
print(multi_measurement([key_qv, value_qv]))
# Yields:
# {('hello', 0.25): 0.5, 
#  ('world', 1.75): 0.5}
\end{minted}
The output shows, that the states of \ttfnt{key\_qv} are now entangled with the states of the values. The quantum type of the \ttfnt{QuantumDictionary} is a custom quantum type (similar to the snippet above). 
An example for how a \ttfnt{QuantumDictionary} can be used for problem solving, is given in Section~\ref{database_oracle}.
\newpage
\subsection{QuantumEnvironments}
\label{quantum_environment}
A \ttfnt{QuantumEnvironment} is a block of code that undergoes a specific mode of compilation. They can be entered using the \ttfnt{with} statement. An important example is the \ttfnt{ConditionEnvironment}. Quantum code within this environment is executed if the condition evaluates to \ttfnt{True}.

\begin{minted}{python}
from qrisp import QuantumFloat, h, control
cond_var = QuantumFloat(5)
h(cond_var)
target = QuantumFloat(5)
with cond_var == 0:
    target += 1
\end{minted}
This snippet computes \ttfnt{cond\_var == 0}, which returns a \ttfnt{QuantumBool}. A perk of this quantum type is that it can be used as a condition environment. Therefore, the incrementation on \ttfnt{target} is executed for the branch where \ttfnt{cond\_var} is zero. \textit{Qrisp} automatically detects if a \ttfnt{QuantumBool} is only used as a \ttfnt{ConditionEnvironment} and uncomputes if this is the case.\\

It is furthermore possible to implement user defined \ttfnt{QuantumEnvironment}s, which can be created by specifying a compilation procedure. The built-in environments are:

\begin{itemize}
    \item \ttfnt{ControlEnvironment}: The compiled quantum code is controlled on a given set of Qubits. 
    Nested control environments are compiled using context informed compilation. Context informed means that for nested control environments the control value evaluation of the inner environment will be compiled controlled on the value of the outer condition. The inner environment's content will however only be controlled on the inner environment's control value. This is a valid optimization because the qubit holding the value on the inner condition will only be in the \ttfnt{True} state if both, the inner and outer conditions are met. This ensures that the code of any given environment requires only a single control qubit compiled\footnote{Compiling a control qubit here means that every quantum gate receives an extra control knob using the technique demonstrated in \cite{Barenco_1995}.} regardless how deep the nesting level is. The \ttfnt{ControlEnvironment} therefore gives a powerful example of how high-level code structure information can be leveraged into low-level compilation performance.
    
    \item \ttfnt{ConditionEnvironment}: Compiled quantum code is executed when a certain condition is met, such as equality of two \ttfnt{QuantumVariable}s. It is also possible to create these environments with custom condition evaluation functions. Similar to the \texttt{ControlEnvironment}, the \ttfnt{ConditionEnvironment} is compiled also in a context informed fashion as described above.
    
    \item \ttfnt{InversionEnvironment}: The compiled quantum code is inverted/daggered.
    \item \ttfnt{ConjugationEnvironment}: Receives a \textit{conjugator} function $f$ together with the corresponding arguments and executes the unitary 
    \begin{align}
        U_{con} = U_f^\dagger V U_f.
    \end{align}
    Where $V$ is the unitary of the code executed within the environment. If called within a \texttt{ControlEnvironment} or \texttt{ConditionEnvironment}, the unitary 
    \begin{align}
    cU_{con} = U_f^\dagger cV U_f
    \end{align}
    is executed, which is much cheaper than controlling all three steps. Similarly to the \texttt{ControlEnvironment}, this is also a compilation technique relying on high-level code structure information.
    
    \item \ttfnt{GateWrapEnvironment}: The compiled quantum code is collected in a single gate object. This environment can be useful for quantum circuit visualization, as it hides the complexity of certain functions. 
    
    \item \ttfnt{GMSEnvironment}: Within this environment only the following gate set is admitable: $\{$Z, RZ, P, CP, CZ, CRZ$\}$. The quantum code in this environment is compiled into an instance of the ion-trap native GMS gate set. These gates allow for entangling multiple qubits within a single step and therefore hold the potential for significant gains in circuit optimization \cite{GMS, GMS_maslov, seidel2021efficient}.
    
\end{itemize}

\subsection{Custom controlled functions}
\label{custom_control}
Controlled function calls are an ubiquitous phenomenon in many quantum algorithms. A generic method for turning an arbitrary quantum circuit containing only CNOT and single qubit gates into its controlled version has been given in \cite{Shende_2006}. The authors approach the problem by introducing controlled versions of arbitrary single qubit gates, which in turn also enable the compilation of CCNOT or Toffoli gates. Even though this procedure is guaranteed to provide a result, in many cases more application specific methods for controlling the circuit are significantly cheaper. Consequently, a variety of papers also describe the controlled version of their circuits \cite{gidney2018, rines2018, beauregard2003}.\\
For developers of circuit level development kits, this usually poses an aggravating extra challenge. If the control Qubit becomes available in some higher stack frame while the custom controlled circuit is hidden below several other function calls, all these functions have to track the control Qubit. This essentially implies that merely creating two different implementations of the circuit under consideration--one controlled and one regular one--is insufficient. Additionally, the function calls between the control Qubit creation and the circuit implementation cannot be reused. Naturally, this prevents the development of modular and future-proof code.\\
Furthermore, in a collaborative setting, developers have to communicate manually that their peers should use the controlled circuit instead of applying the generic control routine to the uncontrolled circuit.\\
Within \textit{Qrisp}, this problem is solved elegantly with the \ttfnt{custom\_control} decorator. Applying this decorator will give a function that, if called within a \ttfnt{ControlEnvironment} or \ttfnt{ConditionEnvironment}, receives the control Qubit as a keyword argument. We demonstrate this behavior by implementing a controlled SWAP operation:
\begin{minted}{python}
from qrisp import *

def regular_swap(a, b):
    cx(a, b)
    cx(b, a)
    cx(a, b)

@custom_control
def custom_controlled_swap(a, b, ctrl = None):

    if ctrl is None:
        env = QuantumEnvironment()
    else:
        env = control(ctrl)
        
    cx(a, b)
    with env:
        cx(b, a)
    cx(a, b)
\end{minted}
Within this snippet we implement two functions: One is custom-controlled and the other one uses the generic control routine. The idea behind the custom controlled version is that in a controlled SWAP, only the middle CX gate has to be controlled because if the control Qubit is off, the outer CX gates cancel each other. To realize this behavior, we first create the quantum environment in which we call the middle CX gate. If the \ttfnt{ctrl} argument is \ttfnt{None}, this implies the function has not been called within a \ttfnt{ControlEnvironment}. As such we perform no controlling operation to the middle CX gate, which is achieved by the base \ttfnt{QuantumEnvironment}, which simply performs it's content without modification. If \ttfnt{ctrl} is not \ttfnt{None}, we know the function has been called within a \ttfnt{ControlEnvironment}, implying we need to turn the middle CX gate into a Toffoli. This is done by calling it within another \ttfnt{ControlEnvironment}. We test the behavior of our function.
\begin{minted}{python}
a = QuantumBool()
b = QuantumBool()
c = QuantumBool()

with control(c):
    regular_swap(a, b)
    barrier([a,b,c])
    custom_controlled_swap(a, b)
\end{minted}
This script generates the following circuit:\\
\begin{center}
\scalebox{1.0}{
\Qcircuit @C=1.0em @R=0.8em @!R { \\
	 	\nghost{{a.0} :  } & \lstick{{a.0} :  } & \ctrl{1} & \targ & \ctrl{1} \barrier[0em]{2} & \qw & \ctrl{1} & \targ & \ctrl{1} & \qw & \qw\\
	 	\nghost{{b.0} :  } & \lstick{{b.0} :  } & \targ & \ctrl{-1} & \targ & \qw & \targ & \ctrl{-1} & \targ & \qw & \qw\\
	 	\nghost{{c.0} :  } & \lstick{{c.0} :  } & \ctrl{-1} & \ctrl{-1} & \ctrl{-1} & \qw & \qw & \ctrl{-1} & \qw & \qw & \qw\\
\\ }}
\end{center}
which shows that the custom controlled function performs the same unitary as the generic one but with less resources required. Regarding the CX count we have: 18 compared to 8.\\
Using the \ttfnt{custom\_control} decorator therefore solves the above-mentioned problems:
\begin{itemize}
    \item If the custom controlled swap is called below a stack of several functions $f_1 .. f_n$, these functions don't need to track the control qubit, implying the same code is valid without modification for the non-controlled call.
    \item Low-level developers can expose controlled versions of their functions, without high-level users ever learning about this extra-layer of complexity, while still benefiting from the additional performance.
\end{itemize}

\begin{figure}[h!]
    \begin{subfigure}[b]{0.31\textwidth}
        \includegraphics[width=\textwidth]{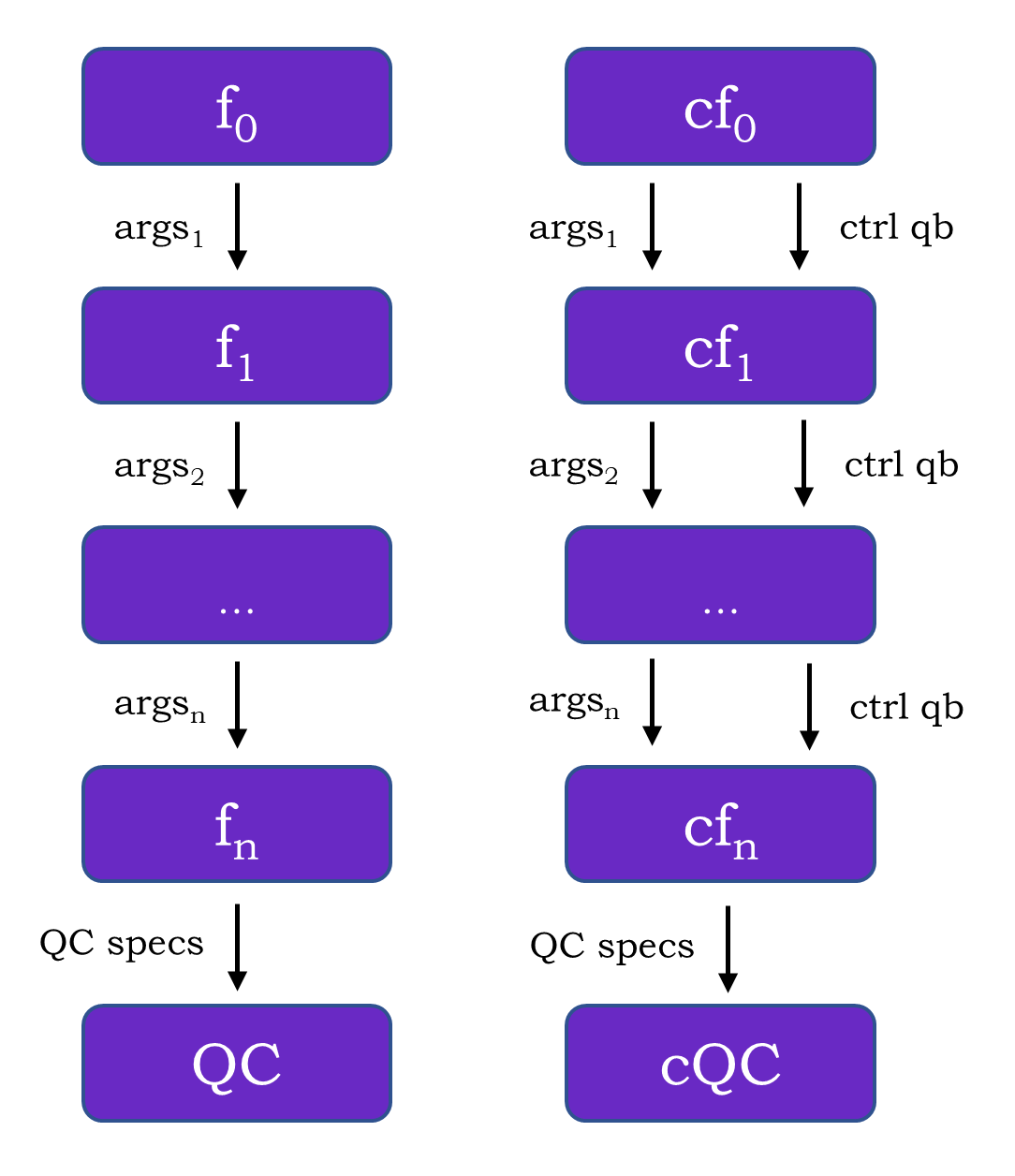}
        \caption{\label{fig:custom_controlled_circuit}}
        \label{fig:subfig1}
    \end{subfigure}
    \hfill
    \begin{subfigure}[b]{0.136\textwidth}
        \includegraphics[width=\textwidth]{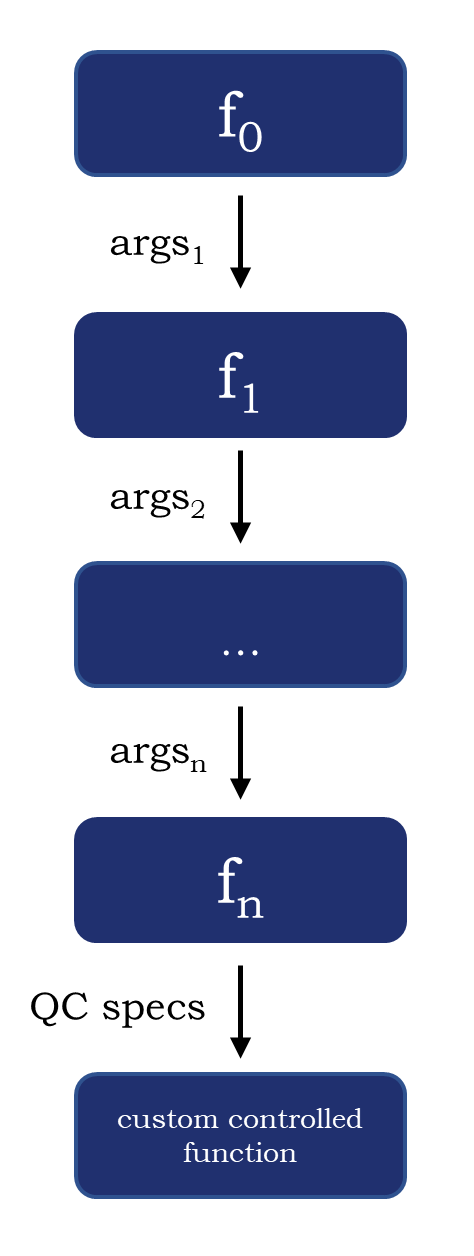}
        \caption{\label{fig:custom_controlled_function}}
        \label{fig:subfig2}
    \end{subfigure}
    \caption{\ref{fig:custom_controlled_circuit} An example call tree as implemented in a circuit construction framework. Displayed is the situation that the developer implements a function $f_0$ that calls a circuit construction procedure hidden below several subroutines $f_1 .. f_n$. To implement a controlled version of $f_0$ using the custom controlled circuit compilation procedure, the developer essentially needs to rewrite every function within the stack in order to accommodate for the appropriate treatment of the control qubit. Users need to be aware that a  custom controlled version $cf_0$ is available. \ref{fig:custom_controlled_function} The same situation but using the \ttfnt{custom\_control} decorator. The tracking of the control qubit is automated, implying only the lowest level needs to be adjusted. Within this approach the custom controlled version of $f_0$ is automatically used (if called within a \ttfnt{ControlEnvironment}), implying users can be oblivious of the custom control procedure.}
    \label{fig:mainfig}
\end{figure}
\section{QuantumCircuit infrastructure}
\label{infrastructure}
This section provides an overview over some of the circuit manipulation infrastructure. Even though \textit{Qrisp}'s goal is to expose a high-level programming interface, refusing to offer  circuit construction features would exclude three decades of research and restrict interoperability. This is why \textit{Qrisp} comes with its own \ttfnt{QuantumCircuit} infrastructure, which is intended as the interface to established programming frameworks and established low-level representations like OpenQASM \cite{Cross2022}.

In order to provide a high degree of compatibility to existing code, we designed \textit{Qrisp}'s \ttfnt{QuantumCircuit} infrastructure in line with the Qiskit \cite{Qiskit} interface. Hereafter we provide some code examples to illustrate how this works in practice. 

Consider an arbitrary piece of Qiskit code constructing a quantum circuit:

\begin{minted}{python}
from qiskit import QuantumCircuit

qc = QuantumCircuit(3, 3)
qc.h(0)
qc.cx(0,1)
qc.cx(1,2)

for i in range(3): qc.measure(i, i)
\end{minted}

In many cases, \textit{Qrisp}'s \ttfnt{QuantumCircuit} uses the same method names as Qiskit's \ttfnt{QuantumCircuit}, i.e., it is sufficient to simply replace the first line to convert the code to a \textit{Qrisp} quantum circuit:

\begin{minted}{python}
from qrisp import QuantumCircuit

qc = QuantumCircuit(3, 3)
qc.h(0)
qc.cx(0,1)
qc.cx(1,2)

for i in range(3): qc.measure(i, i)
\end{minted}

This feature enables a very quick adaption of many existing circuit construction programs, requiring only very few modifications. As \textit{Qrisp} is under active development, some Qiskit features are still missing while many are already included using alternative -- and sometimes better -- algorithms. We discuss two of these cases below.\footnote{A detailed comparison with other circuit creation frameworks will follow in a future publication.}

\subsection{Controlled gate synthesis}
\label{controlled_gate_synthesis}
Controlled gate synthesis is the task of turning an arbitrary (possibly non-elementary) gate into its controlled version. To acquire a controlled gate in Qiskit, we have to call its control method

\begin{minted}{python}
from qiskit.circuit.library import HGate
chGate = HGate().control(3)
\end{minted}

which works in \textit{Qrisp} nearly identically:

\begin{minted}{python}
from qrisp.circuit.library import HGate
chGate = HGate().control(3)
\end{minted}

However, the algorithm employed by \textit{Qrisp} is more efficient, since it makes use of phase tolerant synthesis \cite{Seidel_2023} for the multi-CX gates, which are required in order to perform the controlled gate algorithm presented in~\cite{Barenco_1995} and \cite{Amy_2013}. For example, using the Qiskit code to synthesize a 3 controlled H gate results in a circuit using 58 CNOT gates (using Qiskit 1.1.0). The CNOT count of \textit{Qrisp} for the exact same piece of code is 14, which means a reduction of more than $75\%$. If called from the high-level interface, in some situations even an exponential reduction of resources for controlled gate compilation is possible. Details of this feature are provide in Section~\ref{compilation}.

\subsection{Unitary calculation}
A reliable way of verifying that two circuits indeed performed an identical transformation involves comparing their unitary matrices. In Qiskit the unitary can be acquired by

\begin{minted}{python}
from qiskit_aer import AerSimulator
qc = circuit_construction()
qc.save_unitary()
backend = AerSimulator(method='unitary')
job = backend.run(qc)
result = job.result()
unitary = result.get_unitary(qc).data
\end{minted}

\textit{Qrisp} requires less code to achieve the same:

\begin{minted}{python}
qc = circuit_construction()
unitary = qc.get_unitary()
\end{minted}
Comparison is also very simple in \textit{Qrisp}:

\begin{minted}{python}
qc_1 = circuit_construction_1()
qc_2 = circuit_construction_2()
eq_bool = qc_1.compare_unitary(qc_2, 
                               precision = 10)
\end{minted}
For a substantial number of circuits that we evaluated, \textit{Qrisp}'s algorithm was faster for Qubit counts $>8$. For a multiplication circuit with 14 qubits, Qiskit needed 47 seconds while \textit{Qrisp} only needed 5 seconds.\footnote{A more systematic benchmark will be published in the near future.}

\textit{Qrisp} can also calculate unitaries with abstract parameters (represented through SymPy \cite{sympy} symbols), which is a feature not supported in Qiskit:
\begin{minted}{python}
from qrisp import QuantumCircuit

qc = QuantumCircuit(2)

from sympy import Symbol
phi = Symbol("phi")

qc.cp(phi, 0, 1)

unitary = qc.get_unitary()
\end{minted}
This yields an array of symbolic entries:
\begin{align}
    \left[\begin{matrix}1 & 0 & 0 & 0\\0 & 1 & 0 & 0\\0 & 0 & 1 & 0\\0 & 0 & 0 & e^{ i \phi}\end{matrix}\right]
\end{align}
The treatment of symbolic values is also available for the \ttfnt{.statevector} method of the \ttfnt{QuantumSession} class described in \ref{quantum_session}.
\section{Examples}
\label{examples}
As initially explained, the goal of \textit{Qrisp} is to provide a \textbf{practical tool} for a performant compilation of quantum algorithms. As such, we will now give a variety of examples aiming to demonstrate the purposes of the introduced abstractions.

\subsection{Quantum Phase Estimation}
\label{sec:qpe}
In this example we demonstrate the quantum phase estimation (QPE) algorithm~\cite{kitaev1995, nielsen00}, formulated such that the algorithm itself is a higher-order quantum function. QPE is an important building block in many quantum algorithms \cite{Shor_1997, HHL, montanaro}, and can be implemented in a few lines of \textit{Qrisp} code. The QPE algorithm solves the following problem: Given a unitary $U$ and a quantum state $\ket{\psi}$ which is an eigenvector of $U$:
\begin{align}
   U \ket{\psi} = \text{exp}(i 2 \pi \phi)\ket{\psi}
\end{align}

Applying quantum phase estimation to $U$ and $\ket{\psi}$ returns a quantum register containing an estimate for the value of $\phi$.

\begin{align}
       \text{QPE}_{U} \ket{\psi} \ket{0} = \ket{\psi} \ket{\phi}
\end{align}

Note that this equation assumes, that the second register is sufficiently large to hold $\phi$ with full precision.\\
The following \textit{Qrisp} code implements QPE as a higher order quantum function:

\begin{minted}{python}
from qrisp import QuantumFloat, control, QFT, h    

def QPE(psi, U, precision):
   
    res = QuantumFloat(precision, -precision)

    h(res)

    for i in range(precision):
        with control(res[i]):
            for j in range(2**i):
                U(psi)
   
    return QFT(res, inv = True)
\end{minted}

The first step here is to define \ttfnt{QuantumFloat res} (see Section~\ref{quantum_types}), which will contain the result. The first argument specifies the amount of mantissa Qubits the QuantumFloat should contain, while the second argument specifies the exponent. Having $n$ mantissa qubits and an exponent of $-n$ implies that this \ttfnt{QuantumFloat} can represent the values between 0 and 1 with a granularity of $2^{-n}$. Subsequently, we apply a Hadamard gate to all Qubits of \ttfnt{res} and continue by performing controlled evaluations of $U$. This is achieved by using the \ttfnt{with control(res[i])} statement. This statement enters a \ttfnt{ControlEnvironment} (see Section~\ref{quantum_environment}) such that every quantum operation inside the indented code block will be controlled on the $i$-th qubit of \ttfnt{res}. We finalize the algorithm by performing an inverse Quantum Fourier Transformation of \ttfnt{res}.

Compared to the implementation found in many low-level frameworks, e.g., \cite{Qiskit, cirq_developers_2022_6599601}, the \textit{Qrisp} version comes with the convenience that $U$ can be given as a Python function (instead of a Circuit object) enabling flexible and dynamic evaluations. Additionally, the use of a \ttfnt{ControlEnvironment} can yield significant performance gains if QPE is called within another \ttfnt{ControlEnvironment} due to context-informed compilation (see Section~\ref{quantum_environment}).

Its usage is demonstrated in the following simple example:

\begin{minted}{python}
from qrisp import p, QuantumVariable
import numpy as np

def U(psi):
    phi_1 = 0.5
    phi_2 = 0.125

    p(phi_1*2*np.pi, psi[0])
    p(phi_2*2*np.pi, psi[1])
   
psi = QuantumVariable(2)

h(psi)

res = QPE(psi, U, 3)

print(res.qs.statevector())
#Output: (|00>*|0.0>
#       + |01>*|0.125>
#       + |10>*|0.5>
#       + |11>*|0.625>)/2
\end{minted}

In this code snippet, we define a function \ttfnt{U} which applies a phase gate onto the first two qubits of its input. We then create the \ttfnt{QuantumVariable psi} and bring it into uniform superposition by applying Hadamard gates onto each qubit. Subsequently, we evaluate \ttfnt{QPE} on \ttfnt{U} and \ttfnt{psi} with a precision of 3. Finally, we call the \ttfnt{statevector} method of the \ttfnt{QuantumSession} where \ttfnt{psi} is registered (see Section~\ref{statevector_assessment}) to verify that the result matches our expectations.

\subsection{Quantum Loops}
This example demonstrates that classical programming constructs like loops can be used for programming in \textit{Qrisp}. To enable this, we introduce the \ttfnt{qRange} iterator which takes an integer-valued \ttfnt{QuantumFloat} and performs a loop, depending on the state of that \ttfnt{QuantumFloat}.
\begin{minted}{python}
from qrisp import QuantumFloat, h, qRange

n = QuantumFloat(3)
n.encode(6)
h(n[0])

qf = QuantumFloat(5)

for i in qRange(n):
    qf += i

print(qf)
\end{minted}

In the code above, we first encode the number 6 into \ttfnt{n} and then bring the qubit with significance $2^0$ into a superposition, resulting in the state
\begin{align}
    \ket{\text{\ttfnt{n}}} = \frac{1}{\sqrt{2}} ( \ket{6} + \ket{7} ).
\end{align}
Afterwards, running the \ttfnt{qRange} loop results in 6 additions to \ttfnt{qf} for the $\text{\ttfnt{n}} = 6$ branch and 7 for the other branch. Using Gauss's formula
\begin{align}
\sum_{k = 1}^n k = \frac{n(n+1)}{2},
\end{align}
we find the expected console output:
\begin{align}
\{21.0: 0.5, 28.0: 0.5\}
\end{align}

\vspace{3em}

\subsection{Database oracles}
\label{database_oracle}
In our previous work \cite{Seidel_2023}, we introduced an algorithm for automatic generation of Grover quantum oracles for arbitrary data-structures. We summarize the core ideas briefly:\\
Given is a set of elements $D$ (the database entries), and a database:
\begin{align}
e : \mathbb{N} \rightarrow D, i \rightarrow e(i).
\end{align}
We pick an arbitrary labeling function $l$, that maps elements from $D$ onto bit-strings of length $k$
\begin{align}
l : D \rightarrow \mathbb{F}^k, e \rightarrow l(e).
\end{align}
This labeling function can, for example, be realized by hashing the entries and clipping the binary representation to $k$ bits. The index/label pairs are then converted into a truth table, which is implemented via quantum logic synthesis \cite{soeken2017logic, Meuli2019, Porwik2002, Amy_2019, Seidel_2023}. In the code example below, this step will be performed by loading from a \ttfnt{QuantumDictionary} (see Section~\ref{quantum_dictionary}).\\ We get an operator $U_D$ which acts as:
\begin{align}
    U_D \ket{i}\ket{0} = \ket{i}\ket{l(e(i))}
\end{align}
If we now receive a query object $e_q \in D$ we evaluate the labeling function, tag the corresponding state and uncompute. The query oracle unitary is therefore:
\begin{align}
    O(e_q) = U_D^\dagger T(l(e_q)) U_D
\end{align}
where $T(x) = 1 - 2\ket{x}\bra{x}$ is the operator that phase tags a given bitstring.\\
The \textit{Qrisp} code for specifying the labeling function is:
\begin{minted}{python}
from qrisp import (QuantumDictionary,
auto_uncompute, QuantumFloat, QuantumVariable)
from qrisp.grover import tag_state, grovers_alg

k = 5 #Set label size
def labeling(x):
    # Return clipped bitstring of hash
    return bin(hash(x))[-k:] 
\end{minted}
The next piece of code below defines the query oracle generation function.
\begin{minted}{python}
def db_oracle(db, labeling):

    qd = QuantumDictionary(
        return_type = QuantumVariable(k))
        
    for i in range(len(db)): 
        qd[i] = labeling(db[i])   
        
    @auto_uncompute 
    def query_oracle(index_qf, 
                     query_object = None):
                     
        label_bitstring = labeling(query_object) 
        label_qv = qd[index_qf]
        tag_state({label_qv : label_bitstring})

        return
        
    return query_oracle
\end{minted}
This function first creates a \ttfnt{QuantumDictionary} with return type \ttfnt{QuantumVariable(k)}, which essentially corresponds to bit-strings of length $k$. This dictionary is then filled with the index/label pairs (note that this step has to be done only once to generate arbitrarily many query oracles). Subsequently, we define the query oracle function. This function first evaluates the label of the bitstring, loads the values of the \ttfnt{QuantumDictionary} into the superposition, and finally phase tags the label bitstring. The uncomputation is performed automatically due to the function being decorated with the \ttfnt{auto\_uncompute} decorator (see Section~\ref{uncomputation}).\\
The code below shows an example of how this could be used:

\begin{minted}{python}
# Generate some sample database
lorem_ipsum = "Lorem ipsum dolor ..."
n = 4
data = lorem_ipsum.split(" ")[:2**n]
query_oracle = db_oracle(data, labeling)

# Create index integer
index_qf = QuantumFloat(n)

# Evaluate Grover's algorithm
grovers_alg(index_qf, 
            query_oracle, 
            kwargs = {"query_object" : "dolor"})
            
print(index_qf)
#Yields: {2: 0.9613, 0: 0.0026, 1: 0.0026, ...}
\end{minted}
We conclude that the index of query \ttfnt{"dolor"} is indeed 2.

\subsection{Shor's algorithm}
\label{sec:shor}

\begin{figure*}
\includegraphics[width = 0.95\textwidth]{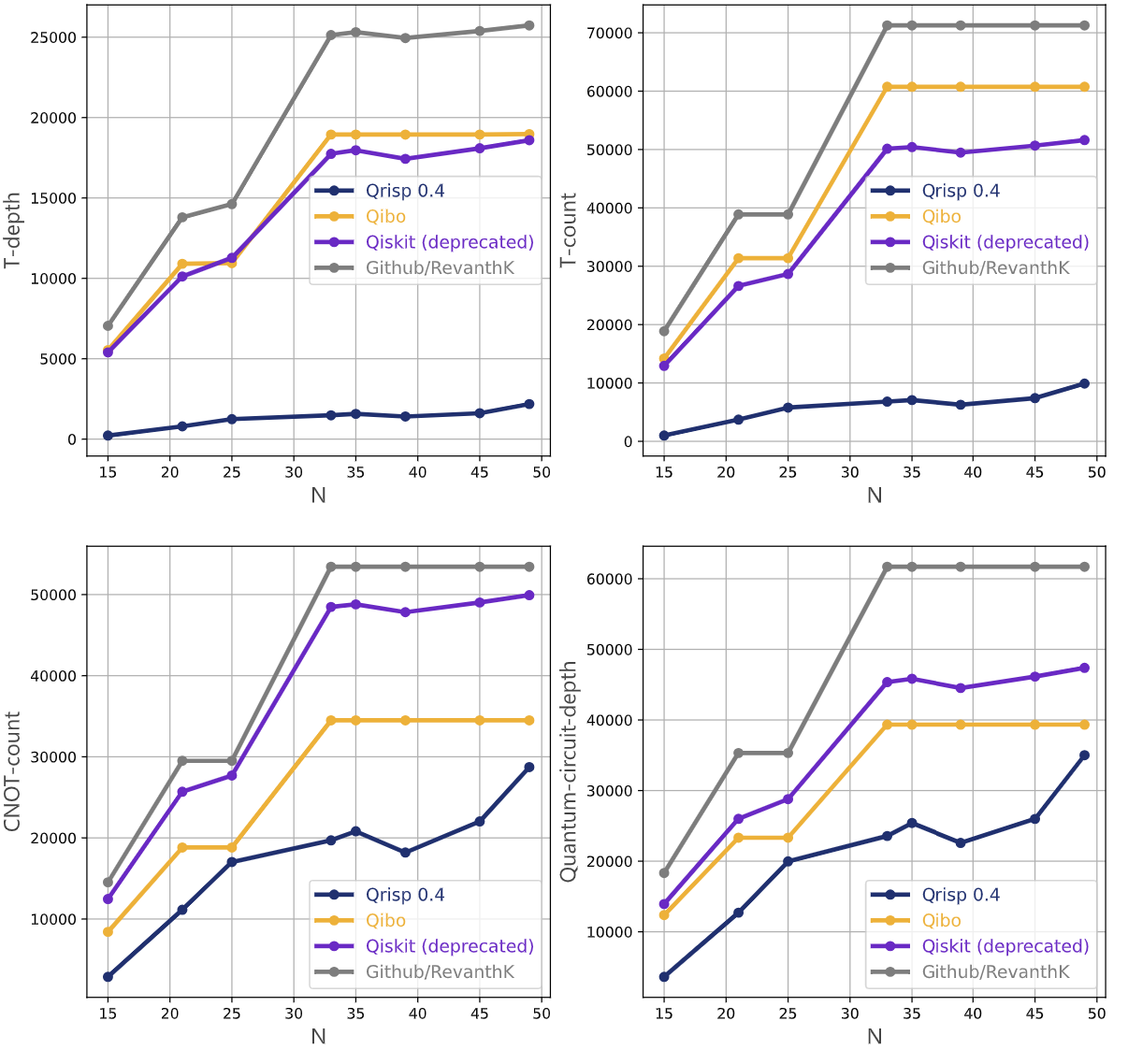}
\caption{\label{fig:shor} Several KPIs of the Shor implementation as described within section \ref{sec:shor}. The utitlized adder is the one described in \cite{gidney2018}. For the quantum fourier transformation we used the technique as described in \cite{park2023} to reduce the amount of arbitrary angle RZ gates. Note that the competing implementations made heavy use of Fourier basis addition, which trades resources such as gate count and circuit depth for arbitrary angle RZ gates. In the Clifford+T gate set, these rotations have to be synthesized using algorithms such as \cite{selinger_2014}. This renders the resource estimation of such algorithms rather difficult since it is unclear to which precision the RZ angle gates need to approximated. In this plot, we take the extremely conservative assumption, that any arbitrary angle gate has T-count/depth 1. A more realistic estimate would give the Qrisp implementation an edge of approximately 3 orders of magnitude.}
\end{figure*}

In the final example we aim to demonstrate how all of the preceding components can work together to compile an implementation of Shor's algorithm \cite{Shor_1997}. This implementation is not only superior in terms of performance\footnote{When compared to the available open-source implementations}, but also almost trivially straightforward, enabling the research and extension of said implementation by scientists from different fields of study. The description here focuses on the implementation of the quantum subroutine. For the classical pre- and post processing we refer to \cite{Shor_1997, lavor2003, beauregard2003}.\\
The central quantum type for implementation of many cryptographic quantum algorithms is the \ttfnt{QuantumModulus}. This data type represents elements of a quotient ring and their processing (also known as modular arithmetic) in superposition. \ttfnt{QuantumModulus} instances can be created by calling the constructor with the desired modulus:
\begin{minted}{python}
from qrisp import QuantumModulus
N = 13
qm = QuantumModulus(N)
qm[:] = 7
qm += 7
print(qm)
# {1: 1.0}
\end{minted}
The obtained result is expected because $7 + 7 = 1 (\text{mod } 13)$.\\
The quantum subroutine of Shor's algorithm is supposed to find the \textit{order}\footnote{The \textit{order} of a modular number $a \in \mathbb{Z}/N\mathbb{Z}$ is an integer $r$ such that $a^r = 1(\text{mod } N)$} of a classically known integer $a$ by performing a quantum phase estimation \cite{kitaev1995} of an operator, which achieves a modular exponentiation of $a$. In turn, this is usually translated into a series of controlled modular in-place multiplications followed by an inverse Quantum Fourier Transform \cite{beauregard2003}. These steps can be described conveniently with the \ttfnt{QuantumModulus} data type:
\begin{minted}{python}
def find_order(a, N):
    qg = QuantumModulus(N)
    qg[:] = 1
    qpe_res = QuantumFloat(2*qg.size+1,
                           -(2*qg.size+1))
    h(qpe_res)
    for i in range(len(qpe_res)):
        with control(qpe_res[i]):
            qg *= a
        a = (a*a)%N
    QFT(qpe_res, inv = True)
    return qpe_res.get_measurement()
\end{minted}
An elaboration of the implementation of quantum phase estimation in \textit{Qrisp} has been given in Section~\ref{sec:qpe}. The relevant distinction here is that this code does not contain the inner loop - instead the group homomorphism property of the in-place multiplication operator $U_x$ can be used to fuse the loop into a singular multiplication with a classically precomputed multiplication factor:
\begin{align}
    (U_x)^n = U_{x^n}
\end{align}
Next, to the striking simplicity of this code snippet, its performance aspects also deserve to be mentioned. A benchmark can be found in Fig.~\ref{fig:shor}. Underneath the in-place multiplication operator there are two more layers of abstraction:

\begin{enumerate}

    \item \textbf{The Montgomery reduction algorithm}, as described in \cite{rines2018}. This layer implements several techniques to perform efficient modular reduction after a non-modular multiplication has been executed. This stands in contrast to implementations like \cite{beauregard2003}, which constructs the modular multiplication by performing modular reduction after \textit{each addition}. While this approach amounts to a more straightforward implementation, the overall amount of non-modular adders is much higher compared to Montgomery reduction. This demonstrates that adopting a structured programming model can lead to measurable performance benefits. This is because more complex techniques can be handled using established techniques from classical software engineering, such as encapsulation and systematic testing/benchmarking of isolated sub-functions.

    \item \textbf{The underlying adder}. Within the Montgomery reduction procedure there are naturally several non-modular adder calls. Non-modular quantum adders are a prevalent topic in literature \cite{draper2000, cuccaro2004, thapliyal2013, gidney2018, wang2023}. However, many of these adders have varying requirements regarding the amount of ancillary qubits necessary for execution. This typically renders them quite inconvenient to implement within manual quantum circuit constructions. The reason being, the programmer has to restructure the entire codebase every time a new adder is implemented in order to amount for the optimal usage of the required ancillae within previous/later steps of the program. Within \textit{Qrisp} this problem is non-existent because the allocation/deallocation mechanism outsources this task to a highly performant automatization. The \textit{Qrisp} implementation therefore allows the user to deploy really arbitrary adders for Montgomery reduction (even user supplied algorithms are possible). Next to some rather simple adders, \textit{Qrisp} also ships with the adder described in \cite{wang2023}, which itself again contains multiple layers of abstraction.

\end{enumerate}
\section{Outlook}
\label{outlook}
As a Python eDSL\footnote{Embedded Domain Specific Language}, \textit{Qrisp} unfortunately inherits not only the merits of the Python ecosystem, but also the drawbacks. One of the biggest hurdles here is the limited execution speed: As compilation happens while Python code is executed, the compilation speed is naturally bottlenecked by the limited execution speed. Algorithms that treat "only" hundreds of qubits are usually compiled within a few seconds, however looking beyond, this issue quickly becomes problematic.\\
Naturally, other areas of computing are also affected by this very problem and came up with a variety of solutions:
\begin{enumerate}
    \item Performing static code analysis of Python source code to compile to a more performant representation such as LLVM. For classical algorithms, this approach has been realized by the Numba framework \cite{numba}. While it is very convenient in many situations, the Python subset that can be Numba compiled is rather limited (it has no class system for instance). To leverage the speed of Numba in practice, programmers usually identify and isolate very specific bottlenecks in their code, since Numba-coding large sections makes the code more difficult to fix and less readable. Since \textit{Qrisp} is specifically designed to improve these aspects, we deem this approach incompatible.

    \item Binding optimized C code to Python functions. This approach leverages the speed of low-level algorithms but still retains the high-level control using the Python abstractions. For quantum, this path has been chosen for TKET and it's subsequent Python binding PyTKET \cite{tket}. We argue however that for quantum compilation this is not feasible because users need gate-level control, implying the compiler has to drop in and out of C after every executed gate.
\end{enumerate}
A third approach has been presented with the Jax framework \cite{jax2018}. Originally conceived for machine learning applications, Jax is able to compile Python code by leveraging a mechanism called \textit{tracing}. Tracing means sending so called \textit{tracers} instead of values through the code, which record the instructions instead of actually executing them. The recordings are stored within an intermediate representation called \textit{Jaxpr}, which is subsequently lowered into a singular C call. Similar to \textit{Qrisp}, the low level implementation of Jaxpr can be readily interchanged, which enables Jax to target a variety of computing architectures such as GPUs, TPUs and CPUs.\\
Apart from established tools, Jax also exposes the possibility of creating new \textit{primitives} that can be connected to user-supplied C-code. Note that this is different from binding Python code to C calls, because Jax combines the given code to produce a singular C call.\\
First approaches to binding quantum compilation to Jax primitives have already been realized in \cite{Catalyst2024Preprint}. Given that, we aim to make \textit{Qrisp} code Jax traceable such that algorithm subroutines (such as adders within Shor's algorithm) have to be traced only once and are subsequently called only "symbolically", i.e. the Python interpreter doesn't have to traverse them again. The flattening to a lower level representation like QIR \cite{nguyen2021} is then outsourced to established classical compilation infrastructure \cite{mlir, lattner2004}.\\
The proposed architecture brings some significant advantages:
\begin{enumerate}
    \item Users can leverage the benefits of high-level abstractions for systematic development of quantum algorithms, yet retaining full control over low-level behavior.
    \item \textit{Qrisp} code achieves the compilation speed of classical languages by leveraging LLVM infrastructure.
    \item It is possible to embed quantum subroutines seamlessly into hybrid algorithms targeting computing architectures such as GPUs, CPUs and TPUs. As such our architecture enables production-ready compilation speed of quantum error correction using a Python interface.
    \item Since the Jaxpr IR is built as a functional programming language, static investigation (such as formal verification) is facilitated.
\end{enumerate}
\section{Conclusion}
\label{conclusion}
\textit{Qrisp} simplifies quantum circuit construction routines to support more complex algorithms and provides a lower entry barrier for developers. We achieve this goal by automating many of the day-to-day tasks of quantum programmers, while also leveraging the Python syntax and ecosystem facilitating more human-readable and efficient implementations of many quantum algorithms. Gate-based programming is supported as well, but can be hidden from the user to a very large extent once a sufficient library of low level functions is established.

An important aspect here is the fact that \textit{Qrisp} code can be modularized effectively. This is because the automated qubit management system (hidden behind \ttfnt{QuantumVariable} (de)allocations) allows several modules to recycle qubit resources for each other without intertwining the code. This feature enables systematic development of code bases, where sub-modules can be readily replaced in case more performant or hardware specific techniques are discovered. Other important benefits are the possibility to perform isolated testing and benchmarking of modules, systematic bug-fixing, and user-friendly documentation. All in all, we expect that using more systematic approaches to quantum programming will heavily benefit scalability and costs.

This claim is supported by the fact that the \textit{Qrisp} implementation of Shor's algorithm outperformed every Open-Source implementation that we could find. Note that this is not attributed to a novel superior approach, but primarily to the integration of several existing techniques, which are already quite complex on their own \cite{rines2018, wang2023}. Because of exponentially difficult software development, such techniques couldn't really be fused together until now with \textit{Qrisp} overcoming said challenges.\\
The very same phenomenon could be observed within another recent work of ours \cite{seidel2024backtracking}, which showcases the quantum backtracking algorithm devised by Ashley Montanaro \cite{montanaro}. To the best of our knowledge, despite it being known for almost a decade, no compilable version of this algorithm is currently available. The implementation within \textit{Qrisp} therefore not only marks the first of it's kind, but also fully separates the backtracking code from the problem specifics, enabling software development for really arbitrary backtracking problem classes.

Even though \textit{Qrisp}'s novelty lies in its high-level programming features, it is important to note, that it also enables seamless programming of lower-level functions. While classical high-level languages like Python quickly integrate new developers, only a tiny fraction of them will ever look at the bit-level implementation of, for instance, the integer comparison operator \ttfnt{<=}.
With \textit{Qrisp}, we observe the opposite: Users who are unfamiliar with low level quantum programming typically start to learn \textit{Qrisp} at a high level of abstraction leveraging the multitude of predefined algorithms, automations etc. The more they progress, the more their interest is drawn towards low level structures - presumably because all levels are written with the same platform, syntax, etc.

With the paper at hand we demonstrate these points through a variety of code snippets. We hope this practical approach can on one hand bring in new developers, while also convincing experts on how a systematic approach to programming can benefit them in their daily work.
\section*{Acknowledgement}
This work was funded by the Federal Ministry for Economic Affairs and Climate Action (German: Bundesministerium für Wirtschaft und Klimaschutz) under the projects with funding numbers 01MQ22005A and 01MQ22007A. The authors are responsible for the content of this publication.

\section*{Code availability}
\textit{Qrisp} is an open-source Python framework.
The source code is available in \href{https://github.com/eclipse-qrisp/Qrisp}{https://github.com/eclipse-qrisp/Qrisp}. A multitude of introductory and advanced tutorials can be found on the webpage \href{https://www.qrisp.eu/general/tutorial}{https://www.qrisp.eu/general/tutorial.html}.

\bibliographystyle{unsrt}
\bibliography{sources}

\end{document}